\RequirePackage{fix-cm}
\documentclass[smallextended]{svjour3}       
\smartqed  
\usepackage{graphicx}
\usepackage{color}
\usepackage{natbib}
\usepackage{amsmath}
\usepackage{rotating}
\journalname{Scientometrics}

\begin{document}

\title{Quantifying the changing role of past publications}

\titlerunning{Evolving co-citations map the changing roles of past papers}        

\author{Katalin Orosz \and Ill\'{e}s J. Farkas \and P\'{e}ter Pollner}


\institute{   K. Orosz \at
              Regional Knowledge Centre\\
              E\"{o}tv\"{o}s Lor\'{a}nd University \\
              H-1117 Budapest, P\'{a}zm\'{a}ny P\'{e}ter s\'{e}t\'{a}ny 1/A
           \and
              I. J. Farkas \at
              MTA-ELTE Statistical and Biological Physics Research Group\\
              Hungarian Academy of Sciences\\
              H-1117 Budapest, P\'{a}zm\'{a}ny P\'{e}ter s\'{e}t\'{a}ny 1/A
           \and
              P. Pollner \at
              MTA-ELTE Statistical and Biological Physics Research Group\\
              Hungarian Academy of Sciences \\
              H-1117 Budapest, P\'{a}zm\'{a}ny P\'{e}ter s\'{e}t\'{a}ny 1/A\\
              Tel.: +36-1-372-2795\\
              Fax: +36-1-372-2757\\
              \email{pollner@angel.elte.hu}
}

\date{Received: date / Accepted: date}

\maketitle

\begin{abstract}
Our current societies increasingly rely on 
electronic repositories of collective knowledge.
An archetype of these databases is the Web of Science (WoS) that stores scientific publications.
In contrast to several other forms of knowledge
-- e.g., Wikipedia articles --
a scientific paper does not change after its ``birth''.
Nonetheless, from the moment a paper is published
it exists within the evolving web of other papers,
thus, its actual meaning to the reader changes.
To track how scientific ideas (represented by groups of scientific papers)
appear and evolve, we apply a novel combination of algorithms
explicitly allowing for papers to change their groups.
We (i) identify the overlapping clusters of the undirected yearly co-citation networks of the WoS (1975-2008)
and (ii) match these yearly clusters (groups) to form group timelines.
After visualizing the longest lived groups of the entire data set we assign topic labels to the groups.
We find that in the entire Web of Science multidisciplinarity is clearly over-represented among cutting edge ideas.
In addition, we provide detailed examples for papers that (i) change their topic labels and (ii) move between groups.

\keywords{Article co-citation network\and Group dynamics\and Tag extraction \and Multidisciplinarity}
\PACS{
01.65.+g 
\and
02.10.Ox   
\and
05.45.Tp  
}
\subclass{
82C41  
\and
91C20  
\and
90B15  
}

\medskip

\noindent {\bfseries JEL} 
B000 
$\cdot$
C810 
$\cdot$
D700 
$\cdot$
O340 
\end{abstract}

\section{Introduction}
\label{intro}

Many current processes generate knowledge in science, technology, medicine and other fields.
Some of these processes are resource-intensive,
for example, biochemistry needs reagents
and experimental subatomic physics needs particle accelerators.
Within each field, and among the ever-increasing number of fields,
the available financial resources need to be distributed properly.
The first step towards a reasonable distribution of financial resources
among the fields of research is the identification of these fields.
The most common solution to this task is to apply the keywords of
publications provided by their authors or assigned by databases.
However, the actual meaning of any fixed keyword appearing on
publications changes over time.
For example, just over the past decade DNA sequencing became a core
aspect of cancer research and cryptography become a core aspect of
mobile communications research.
This implies that the scientific value and societal impact of research
may not be fully accessible by restricting scientometric analyses to
fields identified through fixed keywords only.

In the present paper we propose to follow the fields of science over
time by following which groups of papers are co-cited.
Most importantly, for each publication year (Y) we identify groups in
the network of papers co-cited in year Y. 
In this undirected network the weight ($w$) of a link between
two papers (nodes) indicates that these  
two papers were co-cited $w$ times by papers published in the year Y. 
As an example, for each publication year between 1975 and 2008, we compile
the co-citation network of scientific publications based on the Web of Science.
We find that for several fixed sets of previously published papers
the groups of co-cited papers within these sets change significantly over time.
In other words, the modules of the co-citation network show 
how the scientific community continuously re-evaluates past knowledge
and views it from a continuously changing perspective.
As opposed to defining the fields of science based on keywords
only, this approach can lead to a more accurate identification of
fields and a more precise quantification of impact within each field.
We provide several examples in the paper.

\section{Taking snapshots of the evolution of science and assembling from these snapshots the evolution of topics (fields)}
\label{sci_mapping}

To create a static map of science (a ``snapshot'' of its evolution) we
\begin{itemize}
\item[(i)] retrieved and pre-processed publication data,
\item[(ii)] defined the similarities of publications through co-citation,
\item[(iii)] clustered publications using the co-citation networks to reduce available information, and
\item[(iv)] visualized the obtained map of scientific areas for human analysis.
\end{itemize}

\subsection{Content analysis, bibliographic coupling and co-citation networks}

This section introduces major groups of numerical techniques that have been applied to literature analysis. In the current paper we will be comparing a novel combination of methods to the techniques outlined in this section.

The two major alternatives to author- and keyword-based grouping of scientific content are
to define similarities by (i) {\bf content analysis} (beyond keywords, e.g., title, abstract and main text)
and via (ii) {\bf citation networks}.
A frequently applied {\bf content analysis} technique is co-word analysis,
which allows for discovering the main concepts of any previously selected field
and maps interactions between the pre-selected scientific fields.
In co-word analysis publications (documents)
are labelled with the ``stemmed'' versions of their most characteristic words,
and then labels are connected if they co-occur in at least one document.
Last, in the obtained network of labels concepts are identified as internally densely linked 
groups of nodes and the interactions of a field appear as connections and overlaps among these groups.

While content analysis uses characteristic words of a document,
{\bf citation analysis} uses the references listed in an article's bibliography.
Usually, a citation implies not only that the topics of the citing paper and cited paper are related,
but also that the citing paper makes use of the results of the cited paper.
The first usage of citation analysis dates back to the 1960s.
In 1965, de Solla Price analysed data about the (direct) citations between
scientific papers and identified active research fronts of recent papers in selected fields \citep{price1965}.
Also in the 1960s Kessler introduced a similarity measure called bibliographic coupling \citep{kessler1963}.
Two documents are bibliographically coupled (linked) if there
is at least one other document that they both cite,
and the strength of this connection (the weight of the link)
is the number of documents that they both cite. 
Note that according to bibliographic coupling,
any two papers determine entirely on their own (through their reference lists)
if they are linked and how strongly they are,
and this result remains unchanged over time.
{\bf Co-citation analysis} takes a different approach:
the scientific papers published in a given time interval
decide if and how strongly two earlier papers are linked.
In other words, a bibliographic coupling connection does not change,
whereas a co-citation connection can change.
For example, as scientific activity declines in a given field,
its papers are less frequently cited and also less frequently co-cited.
Thus, a disappearing field of scientific activity gradually disappears also from the co-citation network,
but it remains unchanged in the bibliographic coupling network (with unchanged links and link weights).

Co-citation analysis was suggested in 1973 by
Henry Small \citep{small1973} and Irina Marshakova \citep{marshakova1973}.
Small pointed out that co-citation patterns can quantify 
the relationships between the key ideas of a field with high precision.
Based on this, he suggested applying co-citation analysis to identify scientific fields that emerge quickly, sometimes within a few years.
A technique related to co-citation analysis is co-citation proximity analysis 
where citations appearing in the text closer to each other contribute more to the co-citation weight
of the two cited articles \citep{gipp_beel2009}.
The co-citation network of authors (or journals) is defined similarly to the co-citation network of publications.
For example, two authors are connected in the co-citation network of a publication time window,
if at least one paper published in that time window cites both of them.
In 1981 White and Griffith studied the co-citations of key authors in Information Science \citep{white_griffith1981}.
They found, for example, 
that the extracted modules of authors (based on co-citation profile similarities) were often in accordance with the scientific ``schools'' of this field.

Finally, please note the use of two terms in the literature.
Clusters (communities) of publications and authors co-cited in the past
are often referred to as the ``intellectual base'',
and recent papers joining these clusters are called ``research fronts''.
Here we focus on clusters of past papers, i.e., the intellectual bases.

\subsection{Maps of science}

In many fields of science a common way of understanding measured data is to map the data to a network.
In scientometrics (a field of science) the bibliographic coupling network and the co-citation network list weighted pairwise connections
among publications.
Visualizations of this network are often called maps of science
\citep{chen2004,chen2006,nwb2006,sci2tool2009,vosviewer2010}.
Among the first few examples for mapping science
was a two-piece analysis compiling weighted co-citation networks of scientific papers \citep{small_griffith1974,griffith_etal1974}.
The number of papers co-citing papers A and B became
the weight of the link between the two nodes representing papers A and B.
After discarding links weaker than a selected threshold value
the authors identified major areas of science as connected components of the remaining network.
Then, they analyzed the largest component in more detail by 
Multidimensional Scaling (MDS) and hierarchical clustering
(both numerical techniques use pairwise similarities to visually classify items into subgroups).
They applied also higher link weight thresholds 
(with this change one can locate the ``cores'' of scientific areas).
Later, \citep{small1_etal1985} identified co-citation clusters (areas of science) by
combining data normalization and cluster size dependent clustering
with fractional citation counting and the iterative clustering of clusters.

In addition to scientific publications, scientific advances often form the basis of patents as well.
Patents focus on applicability, and they reference earlier patents with related content.
The co-citation approach has been successfully applied to identify thematic groups
among patents \citep{Laipatent2005} and to predict how technology evolves
in the United States of America \citep{Erdipatent2012}.

\subsection{Assembling the evolution of scientific fields from snapshots}

This section discusses the major methods known in the literature for constructing evolving groups 
of publications. The results in the cited papers should be compared to Figure\,\ref{fig:timelines-size}. 

By the early 1990s, co-citation analysis has become a 
major quantitative technique for mapping the structure
and dynamics of scientific research \citep{braam1991_1,braam1991_2}.
A turning point for these techniques was the introduction
of progressive knowledge domain visualization \citep{chen2004}.
This method (i) derives a sequence of co-citation networks
from a series of equal-length time interval slices,
(ii) merges these slices and
(iii) classifies nodes in this merged network based
on their degrees (neighbor numbers) and node betweenness centralities.
Following this approach \citep{chen2010} introduced a cluster summarization technique
to identify clusters of the co-citation network that correspond to scientific communities.
In addition to identifying clusters,
\citep{klavans_and_boyack_2011} compared the local and global map of Information Science
and set up a model for how science evolves based on data from the 2000-2008 time interval.

\section{Interdisciplinarity and multidisciplinarity}
\label{sec:multi}
Over the past decades many scientific, social and medical problems have become
accessible to scientists trained in fields that routinely use detailed quantitative tools.
For example, physicists designed physiological experiments showing that 
noise produced by a computer can measurably improve human tactile sensation \citep{collins1997}.
Another example is that networks and quantitative sociology have helped to
analyze pairwise friendship connections and map school-wide segregation from them \citep{gonzalez2007}.
For a detailed perspective on the role of research involving multiple fields, see, for example \citep{sinatra2015}.

Generally, \textbf{interdisciplinarity} means that a new discipline arises between previously existing ones,
while \textbf{multidisciplinarity} means that multiple separate disciplines provide their viewpoints on the same problem.
As for a quantitative definition of interdisciplinarity,
\citep{leydesdorff2007}
found in the network of journals (defined based on citation patterns)
that after normalization locally the betweenness centrality of a journal
in this network is a good measure of the level of its interdisciplinarity.
Moreover, \citep{steele_and_stier2000} analyzed the forestry literature and found that
articles drawing information from a diverse set of journals are cited
with greater frequency than articles with a more focused bibliography.
As for the multidisciplinarity of publications,
\citep{levitt_and_thelwall2008} measured
for several topics the frequency of citations to papers published 
in mono- and multidisciplinary journals.
If a journal had a single subject category, then they called it mono-disciplinary,
and if a journal had multiple subject categories, they called it multidisciplinary.
They concluded that multidisciplinary research does not necessarily receive more citations.

Here we use the article co-citation network as a map,
and investigate the dense cores of this network with the time evolving clique percolation method, tCPM \citep{palla_etal2007}.
This algorithm extracts the most dense parts (clusters) of a network
and identifies matching clusters from subsequent (time) steps.
Note that some nodes of the matched clusters can be different.
We find that the members of the evolving dense co-cited article cores 
frequently come from multidisciplinary journals.
In other words, a multidsciplinary paper is more likely to be co-cited for a long time with a stable group of other papers,
and thereby it is more likely to be part of an ``intellectual base''.

Note also that in the Web of Science (WoS)
the category ``multidisciplinary sciences'' on a paper
does not directly indicate that the given paper is multidisciplinary.
In the WoS this category on a paper indicates merely
that the journal where the paper was published is a multidisciplinary journal.
For example, Nature is a multidisciplinary journal, therefore, in the
WoS all publications that appeared in Nature have the category ``multidisciplinary sciences''.
We refer to a paper as multidisciplinary if it does have the WoS category
``multidisciplinary sciences'', regardless of whether its focus is
broad or narrow.
For more detailed analyses of the shortcomings of journal-level categories and
for solutions to the article-level subject classification problem
(based on the analysis of cited literature)
we recommend \citep{glanzel_etal1999a}, \citep{glanzel_etal1999b}, and \citep{glanzel_schubert2003}.
In summary, (i) the categories of papers appearing in highly
specialised journals usually describe these papers' subjects more accurately,
(ii) reclassification can be necessary for papers published in journals 
whose publications are covered by the database selectively, or for papers published in journals that are more general or multidisciplinary.
However,
there are also several studies that make use of the subject categories even at the article level,
for example, \cite{moed_etal1995,porter_rafols2009,albarran_etal2011}.

Our main results related to multidisciplinary research are in Figures\,\ref{fig8} 
and\,\ref{fig9}, and a test of the effect of 
changing the link weight threshold is shown in Figure\,\ref{fig10}.

\section{Data and methods}
\label{data_methods}

We received the following items and a few others from
Thomson Reuters' Web of Science (WoS) for each downloaded paper:
unique ID, publication time stamp, keywords (several types),
and the unique IDs of cited publications.
First, we compiled the yearly co-citation networks of WoS publications (papers).
For example, the nodes of the 1993 co-citation network are those papers 
that were co-cited in 1993 with at least one other paper.
With the same example we note also 
that the nodes of the 1993 co-citation network
are papers that were mostly published before 1993.

For each co-citation network we excluded weak co-citation links. 
Please see Section \ref{clusters} for details about this step.
Next, we identified in each yearly co-citation network
the internally densely connected groups of nodes, i.e., clusters of papers.
The method we applied for identifying these clusters
explicitly allows that the identified clusters overlap.
Last, we joined the yearly co-citation networks into
a single temporal sequence of co-citation networks
containing the life histories of many clusters.

\subsection{Identification of topics and network properties}

\begin{figure*}
\includegraphics[width=1.00\textwidth]{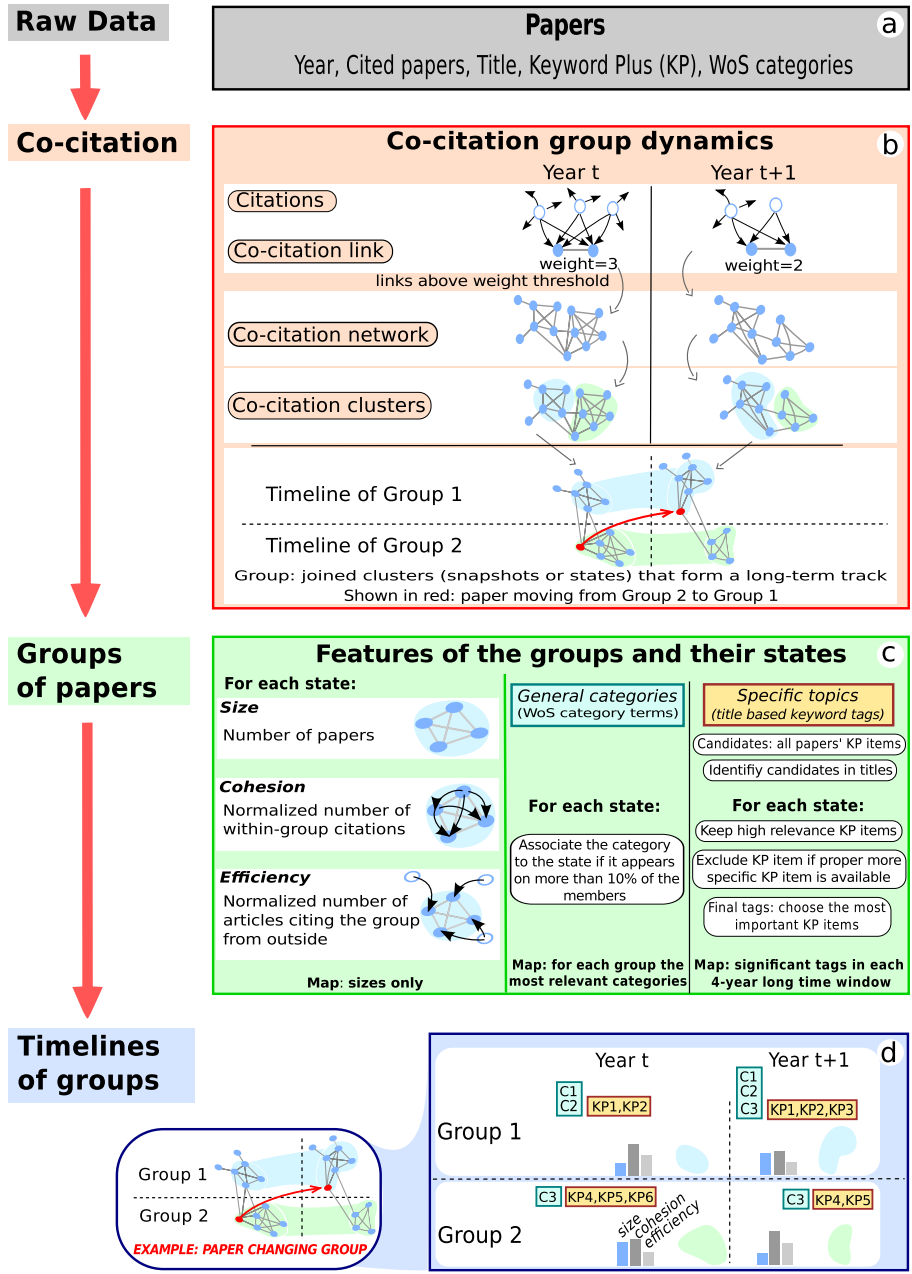}
  \caption{Method scheme.
The method for obtaining the timelines of groups. 
Details are provided in Section \ref{data_methods}.
{\bf (a)} Raw input data.
{\bf (b)} Compilation of yearly co-citation networks, 
their modules (internally densely linked groups of nodes),
and the timelines of the evolving groups of papers.
{\bf (c)} Group properties: 
size, cohesion, efficiency, most important WoS category terms,
more specific topics.
{\bf (d)} Timelines, labels and network-based properties of the groups.
Note that ``older'' citations have a lower contribution to the efficiency.
Group size is the number of nodes in the group.
The cohesion of a subgraph (a set of papers) is defined in Eq.\,(\ref{eq:coh}),
and subgraph (group) efficiency is defined in Eq.\,(\ref{eq:eff}).
}
  \label{fig1}
\end{figure*}

We estimated the specificity of the identified clusters through the WoS categories of cluster members (papers).
Next, we extracted the characteristic topics of each identified cluster
based on (i) the titles of their papers and (ii) a keyword candidate list compiled from the available papers' WoS Keyword Plus tags.
The second method (which is based on WoS Keyword Plus tags) provides a more specific thematic characterization.
After these, we calculated several network properties of the directed article$\rightarrow$article citation network.
We computed group sizes, group cohesion and 
the group's effect on the rest of the scientific community in time.
Finally, we compiled a map visualizing the dynamics of the groups.
This includes the transitions of papers between groups,
changes in the topic composition of groups and group sizes.
Our methods are illustrated in Figure\,\ref{fig1}.

Our co-citation analysis covers the citing years between 1975 and 2008.
As for the directed network of citations,
$17.8$ million articles cite at least one article,
and $16.2$ million articles are cited at least once.
As for co-citations, there are $16.5$ million articles that cite at
least two different articles.
In other words, each of these $16.5$ million articles co-cites at least one pair of articles and contributes to at least one yearly co-citation network.
As for being co-cited, $16.1$ million articles are co-cited with at least one other article.
These $16.1$ million articles are the nodes of the yearly (undirected) co-citation networks.
As a side note, the data set contains $9,481$ nodes with a self-link,
which is a citation link of a publication to itself.
Another special case is when the reference list of article A contains article B more than once.
Between 1975 and 2008 the data set lists $\approx 67,000$ citing articles with this property.
We calculated co-citation weights by including self-citations and repeated out-links.
After obtaining all co-citation weights, we excluded the co-citation of any article with itself.

\begin{figure*}[!ht]
  \begin{center}
    \includegraphics[width=0.7\textwidth]{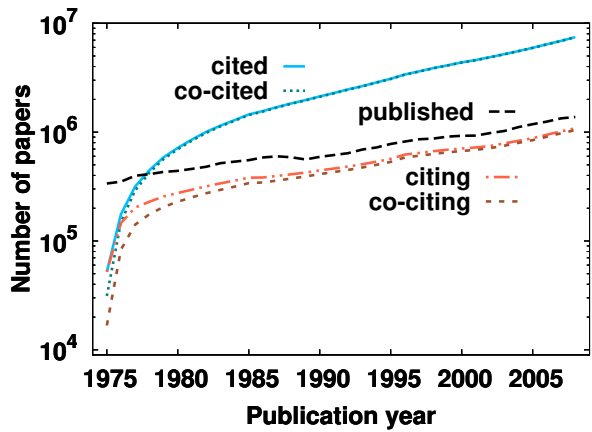}
    \caption{Yearly publication and citation numbers from the Web of Science:
      the total number of papers published in a year (published),
      the number of papers published in a given year and citing at least one paper (citing) 
      or at least two different papers (co-citing),
      the number of papers cited by papers published in the given year (cited),
      and the number of papers cited together with at least one other
      paper by papers published in the given publication year (co-cited).
}
    \label{fig2}
  \end{center}
\end{figure*}

\begin{figure*}[ht]
  \includegraphics[width=1.1\textwidth]{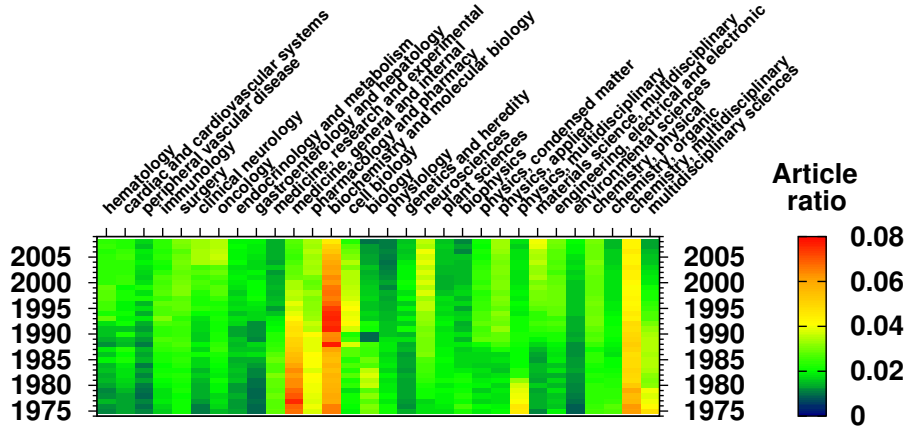}
  \caption{Yearly usage of categories in the Web of Science.
    For each publication year ($t$) and each category ($c$) we show the article ratio, $N_{t,c}/N_t$,
    where $N_t$ is the total number of papers published in year $t$ and $N_{t,c}$ is the number of papers
    that used the category $c$ and were published in year $t$.
    Only categories appearing on more than $2\%$ of 
    all papers in at least one year between 1975 and 2008 are shown.
    Along the horizontal axis categories falling under related fields
    are placed next to each other.
}
  \label{fig3}
\end{figure*}

Regarding the number of publications per year, \citep{szanto_etal2014} found that 
between 1970 and 2010 in the Web of Science and several other databases
the number of papers doubled approximately every 20 years.
Figure\,\ref{fig2} shows that the number of published papers, the number of cited,
co-cited, citing and co-citing papers also grow approximately exponentially.
As for categories, papers published between 1975 and 2008 are categorized
by the Web of Science into $228$ different journal-based categories.
Figure\,\ref{fig3} shows how the number of publications using the most frequently appearing categories changes over time.
For example, the categories \textit{"biochemistry and molecular biology"},
\textit{"medicine, general and internal"},
\textit{"chemistry, multidisciplinary"}
are the most frequently used ones in several years.

\subsection{Clusters of the yearly co-citation networks as states (``snapshots'') of scientific fields}
\label{clusters}

To identify the fields of science as clusters of the co-citation network,
we first compiled for each year the network of papers co-cited in that year. 
In the co-citation network of year $t$
two papers, A and B, are connected by a weighted undirected 
co-citation link, if at least one paper published in year $t$
has both A and B in the list of its references.
For example, in the 1990 co-citation network 
the weight of the link connecting the papers A and B
is the number of those 1990 publications that cite both A and B.
To exclude weak co-citation connections,
we applied edge weight thresholds: links weaker than the threshold
were discarded before the analysis, while weights above the threshold
were set to 1.
For each yearly co-citation network we defined the threshold such
that the group detection method, CPM, could identify the broadest
possible distribution of group sizes.
According to \citep{palla_etal2005}, this choice
of the link weight threshold can provide the most informative clusters (also called: groups or modules).
On a more technical note, setting the link weight threshold
parameter to its optimal value
allows clusters of all sizes to appear, because
the link density of the network is close to the value
at which all nodes are densely linked to same large cluster.
At a higher than optimal
link weight threshold most nodes have no connections
and remain isolated, while at a lower than optimal link weight threshold most nodes
are members of a single large cluster.
Neither of these two extremes is as informative as 
the optimal link weight threshold value that provides
a broad distribution of cluster sizes (clusters of all sizes are present).

As explained above,
in each of the yearly co-citation networks
we identified the dense overlapping groups of co-cited papers
with the Clique Percolation Method (CPM),
which uses undirected links without weights as input.
The CPM identifies overlapping, internally densely linked
clusters of nodes in networks \citep{palla_etal2005},
while CFinder is a software that runs the CPM.
As a technical comment, we mention that the CPM identifies maximal
chain(s) of overlapping complete subgraphs ($k$-cliques),
which are called $k$-clique percolation clusters.
Here we set the clique size parameter to $k=4$.
According to \citep{palla_etal2005},
the optimal value of the clique size parameter, $k$, 
can be selected with a method similar to the selection
of the optimal link weight threshold (see above).
We note also that several of the co-citation networks contain dense
parts (subgraphs) in which exact clique finding for the Clique
Percolation Method is not possible within reasonable computational
time.
In these cases we applied the built-in approximate clique finding
option of CFinder.

\subsection{Joining yearly co-citation clusters (snapshots of scientific fields) into long-term tracks of evolving fields}
\label{join}

The previous section explained how we identified overlapping clusters in the yearly co-citation networks.
In this section the identified clusters are treated as snapshots of the evolving fields of science,
and these snapshots are joined into histories (timelines) of scientific fields with the network module 
joining method of \citep{palla_etal2007}.
First, for any two subsequent years we constructed a network that is
the union of the co-citation networks of these two years.
Next, we identified with CFinder the modules of this merged network.
Note that the merged network fully contains both of the two initial networks,
therefore, each module of the two yearly networks is fully
contained by a module of the merged network.
Consequently, one can identify how modules evolve between two (adjacent)
yearly networks by analyzing -- for each module of the merged network --
which modules of the two yearly networks it contains.
To achieve this we used the notion of relative overlap 
(also called: Jaccard correlation) between any two modules.
If module A from the first yearly network and module B from the
second yearly network share $I_{AB}$ nodes (intersection of A and B), 
and they have a total of $U_{AB}$ nodes (union of A and B),
then their relative overlap is $I_{AB}/U_{AB}$.

In the merged network
we calculated the relative node overlap of all possible A-B module
pairs, where module A is from the earlier yearly network and 
module B is from the later yearly network.
Finally, for any given A-B pair of modules we used the following method to
decide whether module B is a continuation in time (i.e., a later state) of module A.
In each module of the merged network
we took the module pair with the highest relative overlap
and matched these two.
Next, from the remaining (so far unassigned) modules of the initial two networks
we took the module pair with the highest relative overlap and matched these two.
We continued with this process as long as both initial networks had at least one unassigned module.
We joined only module pairs with at least one overlapping node, 
in other words, the relative overlap had to be positive.

\begin{sidewaysfigure}[htbp]
  \vspace*{0.6\textwidth}
  \includegraphics[width=1\textwidth]{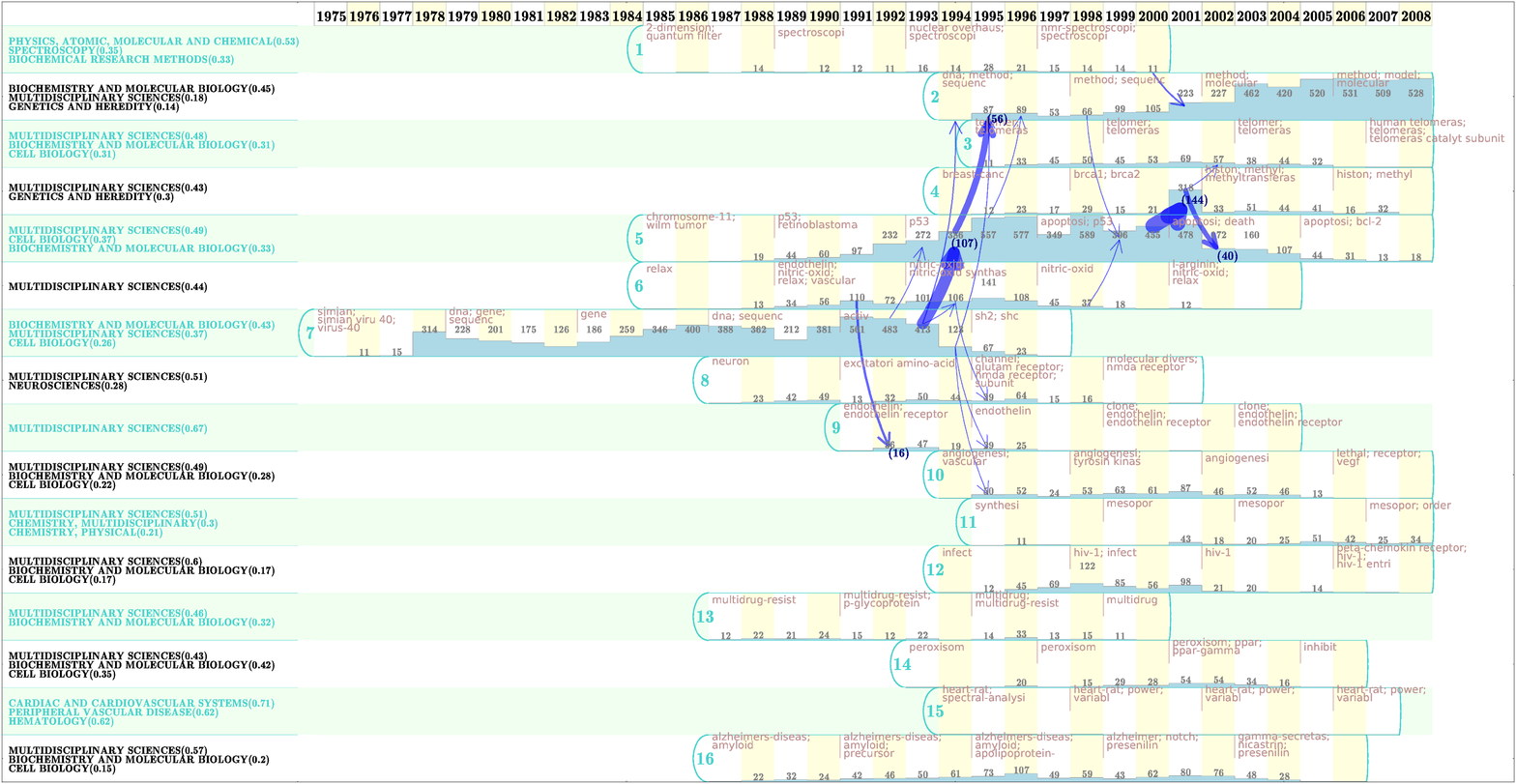}
  \caption{Groups of co-cited articles and articles moving between
    these groups.
    For each year 
    (i) the co-citation network of articles (papers) was 
    defined using Web of Science citation (WoS) data and
    (ii) the dense overlapping groups of co-cited articles were
    identified with CFinder \citep{palla_etal2005}.
    The article groups identified for individual years were 
    joined to form multi-year paths of groups with the method 
    of \citep{palla_etal2007}, see also text.
    Each row corresponds to a group, and
    the height of the rectangle representing the group
    is proportional to the number of papers in the group.
    A transition is indicated with an arrow pointing from the source
    group in a year to the target group (where the papers move) 
    in the next year.
    For each group its most relevant WoS categories are listed at the
    start of the row together with the relevance of the category
    (average portion of the group's papers using that category) in parentheses.
    For each 4-year time window the most significant topics are highlighted.
    Only groups with a lifespan of at least 14 years are shown.
    Group and transition sizes are shown only above size 10.
}
  \label{fig:timelines-size}
\end{sidewaysfigure}

The module pairing process above may be also viewed as the
identification of ``paths'' showing how scientific fields evolve.
Note also that the technique described above includes the possibility
that a module appears (it is ``born''), disappears (it ``dies''),
splits or merges with another module.
These and further details are shown in Figures 1e and 1f of \citep{palla_etal2007}.
The map in Figure\,\ref{fig:timelines-size} shows the histories of the groups
identified with the above method and the transitions among these groups.
For example, an arrow pointing from group $G1$ at time $t$ to group $G2$ at time
$t+1$ means that some papers moved from $G1$ to $G2$ in this time step.

\subsection{Structure of the groups of co-cited publications}
\label{subsec:struct}

This section explains how we investigated whether two papers that are cited together cite each other.
We listed all citations pointing from a paper (citing) to another
paper (cited) that is in the same yearly co-citation cluster.
Note that self-citations were excluded here.
Next, we computed the cohesion, $\kappa(G(t))$, of each group $G$ for each year, $t$.
We denote by $|G(t)|$ the number of papers in group $G$ 
at time $t$ and by $a_{i,j}=1$ a direct citation (directed link)
pointing from paper (node) $i$ to paper (node) $j$:

\begin{equation}
  \kappa(G(t)) =\frac{  \sum\limits_{i,j \in G(t)} a_{i,j}} { |G(t)| \, ( |G(t)| - 1 )},
  \label{eq:coh}
\end{equation}

\noindent
The cohesion of a subgraph in a directed network can vary between 0 and 1.
Note that $\kappa=0$, if and only if the investigated subgraph is empty.
In the case of the network of citations $\kappa=0$ means that there 
is no citation that connects two nodes of the analyzed set of publications.
On the other hand, $\kappa=1$ means that each node of the subgraph
has a directed link pointing to every other node of the subgraph.
For subgraphs in the network of citations 
this translates to each of the $n$ papers citing all $n-1$ other papers.
In other words, if $\kappa=1$, then
any two papers have a citation link in both directions between them.
Note that mutual citation between two articles is rare, because
a citing article usually appears significantly later than the article
it cites.
However, the data set does contain pairs of mutually citing papers.
The total number of citations (directed links) between the articles in our
dataset is over $305$ million. 
Note also that there are around $110$ thousand (A,B) pairs of papers 
for which A cites B and B cites A as well.
Despite their low overall ratio ($0.1\%$), bidirectional links 
are often enriched in specific publication types
(e.g., articles in the same conference proceedings booklet often cite each other).
This is why we take into account the possibility of mutual links
 in the normalization factor of $\kappa(G(t))$.

\subsection{The influence of the identified groups of co-cited papers on science}
\label{subsec:influ}

We computed the efficiency, $\varepsilon(G(t))$, of each group, $G$,
for each year, $t$, to measure how many papers from outside $G(t)$ 
cite the papers of $G(t)$ (the group of articles co-cited in year $t$),
see Eq. (\ref{eq:eff}).
In the definition below, $t_i\leq t$ is the publication 
year of paper $i$ and $\lambda = 0.23$ is a constant.
We selected this particular value of $\lambda$
to set the relative contribution of a paper published 10 years before 
the co-citation year to $\exp(-10\,\lambda)=0.1$.
In Eq. (\ref{eq:eff}) the summation runs for all papers, $i$,
that fulfil all of the following conditions: they
(a) are not in $G(t)$, 
(b) cite at least one paper of $G(t)$, and
(c) were published in year $t_i$, for which $t_i\leq t$:

\begin{equation}
  \varepsilon (G(t))=\sum_i \frac{e^{- \lambda (t-t_i)}} {|G(t)|},
  \label{eq:eff}
\end{equation}

\noindent
Recall that $G(t)$ is a group of papers with
dense pairwise co-citations received in year $t$,
thus, the papers in $G(t)$ were all published 
not later than the year $t$.
Note also that the contribution of citations to
$\varepsilon(G(t))$ is
exponentially decreasing with the time difference $(t-t_i)$,
thus, the strongest contribution to $\varepsilon(G(t))$ 
is provided by citing papers
published in year $t$ or short before year $t$.
Consequently, the efficiency, $\varepsilon(G(t))$,
quantifies the effect of the group of publications, $G(t)$,
on science in year $t$ and short before year $t$.

\subsection{Thematic analysis of groups}

The previous sections introduced yearly co-citation networks and the groups of co-cited publications.
The current section explains how we defined topics for each group of publications.
In Sections \ref{subsubsec:journal} and \ref{subsubsec:kwp_and_title}
we explain two other methods for listing the categories of the groups of papers.

\subsubsection{Journal-based paper categories}
\label{subsubsec:journal}

First, we assigned to each paper the Web of Science categories of its journal.
(Regarding papers with the category \textit{``multidisciplinary sciences''},
please see also the second part of Section\,\ref{sec:multi} above.)
Next, for each group we determined its top three (most relevant) categories.
We considered the entire path of the group over the
analyzed year range and we ranked those categories
that are present on more than $10\%$ of the group's 
articles in more than $70\%$ of all years of the group.
To rank the categories of a group we performed the following steps.
First, for each year ($t$) we computed the total number of papers in the group in that year ($N_{t}$)
and for each category ($c$) the number of the group's papers using that category in year $t$ ($N_{t,c}$).
Second, we averaged the ratio $N_{t,c}/N_{t}$ over all years in which the group exists.
Third, we assigned a lower rank (higher significance) to a category if its averaged ratio was higher.

Interestingly, the presence of the category \textit{``multidisciplinary sciences''}
among the top categories indicates the abundance of
articles published in ``multidisciplinary'' journals,
and cannot be reliably applied to quantify article-level multidisciplinarity.
Moreover, since categories are assigned to journals (and not individual articles) in the Web of
Science, tracking subjects with a resolution significantly higher than
the journal level would require further information at the level of articles.

\subsubsection{Paper categories based on Keyword Plus and titles}
\label{subsubsec:kwp_and_title}

This section explains a combination of previously published methods for identifying the most relevant topics of paper groups.
We (i) extract possible topics from paper titles using the Tf-Idf technique (Term frequency - Inverse document frequency) \citep{salton_and_buckley1988}
and then (ii) select the most relevant of these possible topics with the rCUR dimension reduction method
\citep{bodor_etal2012}.

Among the Web of Science (WoS) data fields available to us,
Keyword Plus\footnote{KeyWords Plus\textregistered\, are ``index terms
created by Thomson Reuters from significant,
frequently occurring words in the titles of an article's cited references.''}
tags are the most appropriate for describing topics with a resolution at the article level. 
Note, however, that of all $27.8$ million papers in the
subset of the Web of Science available to us, $14.8$ million papers ($53\%$) have no KeyWord Plus.
Therefore, we collected all available Keyword Plus tags (KWP),
treated these KWP as keyword candidates and searched for these candidates
in the titles of all group member papers.
Note also that keywords can be expressions, thus, both keyword candidates and title words had to be stemmed (normalized).
We normalized each word with the Porter stemmer \citep{porter1980} (implemented by the NLTK module in Python).

After listing the set of keywords of a selected yearly co-citation
group, we kept only the most relevant of these keywords.
The relevance score we applied is the Tf-Idf score \citep{salton_and_buckley1988}.
The Tf-Idf score is high if the given keyword occurs in many paper titles
in the given yearly co-citation group,
and it is low if the keyword occurs in many of the groups of the 
selected year:

\begin{equation}
  \text{Tf-Idf}(kw,G(t))=\text{Tf}(kw,G(t)) * \log \left( \frac{N_g(t)} {n_g(kw,t)} \right) ,
  \label{eq:tf_idf}
\end{equation}

\noindent
where $\text{Tf}(kw,G(t))$ is the number of articles in $G(t)$ that have the
keyword $kw$ in their titles, $N_g(t)$ is the number of groups
in year $t$, and $n_g(kw,t)$ is the number of those groups in year $t$ in which $kw$ appears. 
After computing the Tf-Idf score for the keywords of each group
we applied two absolute thresholds to keep only the most relevant
keywords.
First, we excluded keywords present in only one article title in the group (in the particular year).
Second, we selected the 10 keywords with the highest Tf-Idf scores.
Moreover, to focus on keywords that are specific enough,
we excluded a keyword also if it is part of another keyword
that has a higher or equal Tf-Idf.
For example, one of the article modules in the 1976 co-citation network consists
of four articles with the following titles:
\begin{itemize}
\item[1.]"Interspersion of repetitive and nonrepetitive \textit{dna sequences} in drosophila-melanogaster genome"\\
\item[2.]"\textit{Dna sequence} organization in genomes of 5 marine invertebrates"\\
\item[3.]"Structural genes adjacent to interspersed repetitive \textit{dna sequences}"\\
\item[4.]"Comparative aspects of \textit{dna} organization in metazoa"\\
\end{itemize}
For this module the following three topic tags have the highest Tf-Idf scores
(Tf-Idf scores are in parentheses): 
\textit{dna sequenc} (11.39); \textit{dna} (8.74); \textit{sequenc} (7.23).
From these three we kept only \textit{dna sequenc}. 
We excluded \textit{dna}, because \textit{dna sequenc} contains it
and \textit{dna} has a lower Tf-Idf score than \textit{dna sequenc}.
Similarly, the topic tag \textit{sequenc} was excluded as well.

Next, we selected for each group of papers the three most relevant keywords with the rCUR algorithm.
For a concise description of the rCUR method let us first select a year and
consider a group of papers that are strongly co-cited in that year.
For this group of papers the rCUR method provides a matrix, $M$,
in which a row corresponds to a keyword of the group
and a column corresponds to an article of the group.
Moreover, the matrix element $M_{i,j}$ is $1$, if and only if the $i$-th 
keyword is present on the $j$-th paper, otherwise it is $0$. 
(Only articles that have at least one of the keywords are included.)

The rCUR method approximates the input data matrix with a small number
of its rows.
This means that a few of the keywords are selected as representative keywords.
To achieve this, first, the sufficient number of top singular values
(and their vectors) is selected: the sum of selected singular values
should exceed $80\%$ of the sum of all singular values.
Second, the leverage score (explained below) for a row of the input data matrix is computed.
(Recall that a row of the input data matrix corresponds to a keyword of the group of papers.)
The leverage score for a selected row of the input data matrix is, up to scaling,
equal to the diagonal element in the same row of another matrix
that projects all vectors onto the subspace of the selected singular vectors.
Last, the rows with the highest leverage scores are selected.

Compared to other data reduction techniques
the key advantage of the rCUR method is the following.
Instead of selecting a linear combination of keywords as the most
relevant one, rCUR selects directly some of the actual keywords.
From the rCUR package we applied the top.scores method, 
and accepted the three rows (keywords) with the highest leverage scores.
We note also that if the number of keyword candidates is low
or the articles of the selected group have very similar keywords,
then the rCUR method is not applicable.
In the first case (few keyword candidates) we selected all keyword
candidates as keywords,
while in the second case (similar keyword occurrence profiles)
we selected the top three keywords as ranked by the Tf-Idf scores.
Finally, for groups existing over at least 14 years,
we compressed the extracted tag information in 4-year time windows
and visualized the extracted tags of the groups in these time
intervals (see Figure\,\ref{fig:timelines-size}).
In this compression step we consider a tag (characteristic topic) of
a group significant if it appears in more than $50\%$ of
the time steps of the range.
If, however, we find no tag above $50\%$, then $50\%$ is accepted.

The consistency of the extracted categories (explained in Section\,\ref{subsubsec:journal})
and the topic tags (Section\,\ref{subsubsec:kwp_and_title})
is illustrated by the groups G5 and G8 in Figure\,\ref{fig:timelines-size}.
For group G5 the top three categories are \textit{``multidisciplinary sciences''},
\textit{``cell biology''}, and \textit{``biochemistry and molecular biology''}.
Group G5 existed between 1985 and 2008, that is, in 24 consecutive years.
During this time $\sim 2,000$ articles were group members in at least one year.
The following topic tags are associated with the group in at least 3 time steps (years):
apoptosi, p53, death, chromosome-11, wilm
tumor, retinoblastoma, kinas, gene, cell-cycl, bcl-2.
Note that topic tags are stemmed expressions (made up of one or more words), for example, 
a stemmed expression is ``kinas'', which replaces a long list of terms containing ``kinase'', ``kinases'' and others.
For Group G8 the top categories are
\textit{``multidisciplinary sciences''} and \textit{``neurosciences''} (Figure\,\ref{fig:timelines-size}).
The group existed for 15 years (between 1987 and 2001)
and contained $\sim 150$ articles during this period.
Its topic tags occurring in at least 4 years (i.e., 4 time steps)
are: nmda receptor, glutam receptor, nmda, neuron, excitatori amino-acid, channel.
As illustrated by these two examples, the categories of article groups
are indeed informative about a group's thematic location in science.

\section{Results}
\label{results}
\begin{figure*}[b]
  \includegraphics[width=1.00\textwidth]{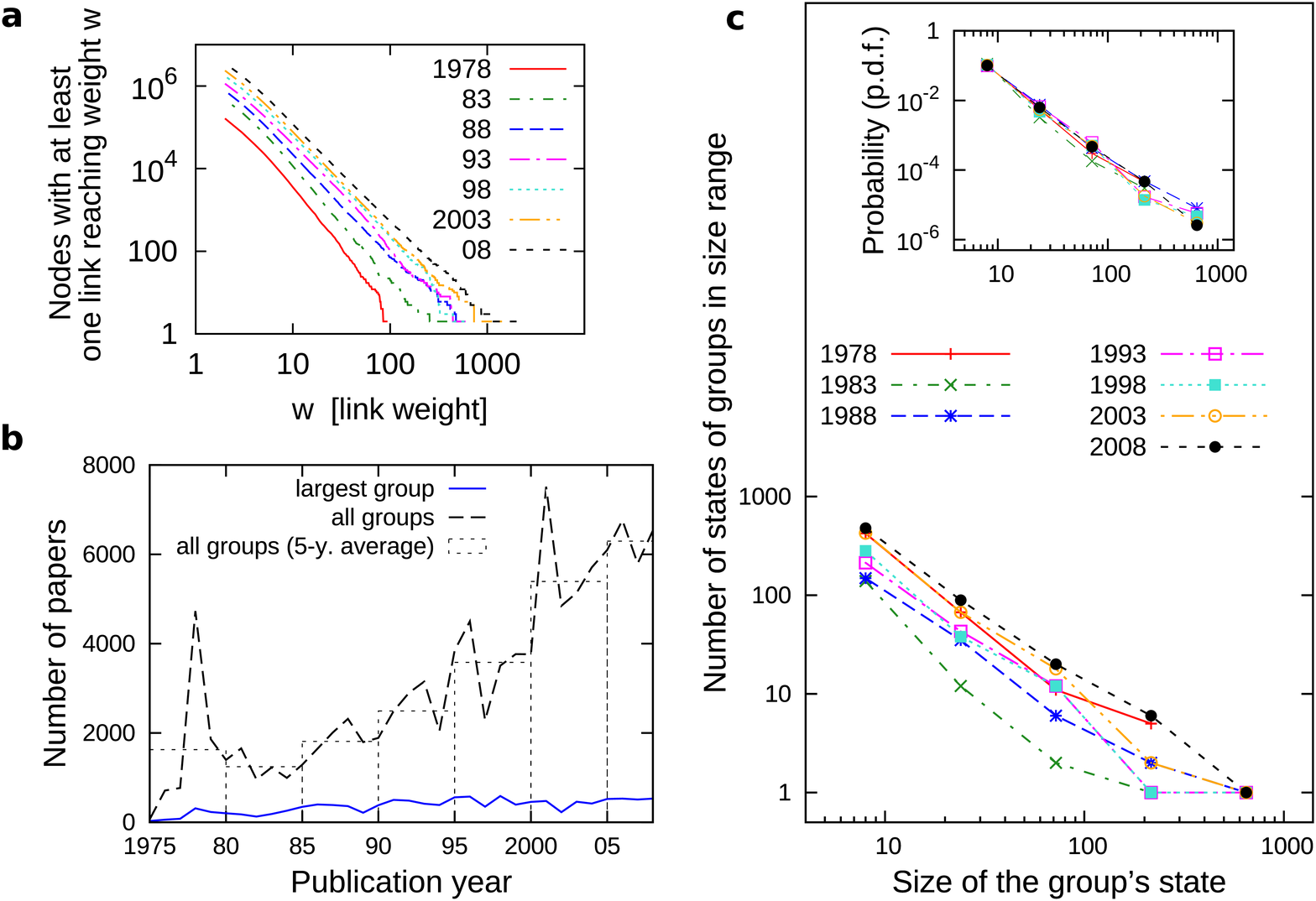}
  \caption{
    Co-citation network and group properties in the analyzed
    time range for every fifth year.
    (a) Number of papers in the co-citation network 
    (i.e., number of nodes with an edge) depending on the
    edge weight threshold.
    (b) Number of articles in the groups (clustered articles) and the size of the
    largest group in each year when applying the optimized link weight
    threshold.
    (c) Number of the states of groups (``snapshots'' of group histories) with various sizes at the
    optimized weight thresholds (aggregated in logarithmic size ranges).
    The corresponding probability density is shown in the inset.
    Note in panels (a) and (c) that despite the changing number of publications
    the overall distribution of co-citation link weights and group sizes
    remains continuously broad. 
    In other words: it is not possible to draw a sharp line
    between small and large co-citation link weights or 
    between small and large groups.
    Observe also in panel (b) that -- from the 1980s to the 2000s --
    compared to the sharp growth of the total number of
    papers of all co-citation groups,
    the largest group grows more slowly.
    Also, the size distribution of the states of groups is unchanged.}
  \label{fig4}
\end{figure*}

Scientific knowledge is evolving mainly through novel results.
The history of this evolution is tracked by scientific publications.
Each paper records the addition of a new item to our shared knowledge.
Until now scientific publications have been considered to be static,
because after they appear they do not change.
Here we show that even though the content of a published paper is ``frozen'', its role (its meaning) often changes over time.
In our results we build our observations mainly on the statistical
analysis of the constructed co-citation networks and co-citation
article group properties with special attention to changes over time.
We observe that over time the yearly co-citation network grows approximately exponentially.
We observe also that if we remove co-citation links that are weaker than a selected link weight (co-citation number) threshold,
then the number of nodes remaining in the network decreases as a power law function of the applied link weight threshold.
This latter result holds for all analyzed years (see Figure\,\ref{fig4}a),
thus, in all publication years the strongest co-citation activity is focused on a small group of highly cited past papers.
These form the (previously mentioned) intellectual base of science.

Figure\,\ref{fig4}b shows 
the annual number of papers clustered into groups
and the size of the largest group obtained from the co-citation networks 
(with link weight thresholds adjusted as explained in Section\,\ref{clusters}).
Interestingly, in contrast to the exponential growth of the number of publications,
the number of articles in the dense core (members of at least one co-citation group)
grows only slowly.
Note also that the sizes of the identified groups differ largely.
This is shown in Figure\,\ref{fig4}c with the corresponding probability
density function (p.d.f.) in the inset.
The power law shape of the p.d.f. of group sizes indicates 
that the typical group is small (about a dozen of articles),
however, a few groups are very large, of the order of the amount of nodes in the small groups combined.

In summary, Figure\,\ref{fig4} quantifies clearly a qualitative
concept that has not yet been quantified at this scale with such
precision.
This concept is that in scientific research most innovations last only
for a few years, while a few may survive for over a decade before
being integrated into other fields.
By comparing this figure to Figure\,\ref{fig:timelines-size} one
may note that in Figure\,\ref{fig:timelines-size} small transitions of
papers indicate rearrangements among fields of research, while large
transitions usually indicate the assimilation of a field into others.

Most importantly, Figure\,\ref{fig4} demonstrates the knowledge
filtering and compressing role of the global publication system in
science: the ``production'' of science increases exponentially,
however, the core knowledge accumulates at a much slower rate.

\subsection{Groups of co-cited papers frequently overlap}
\label{cocitmod}

\begin{figure*}[b]
  \includegraphics[width=1.00\textwidth]{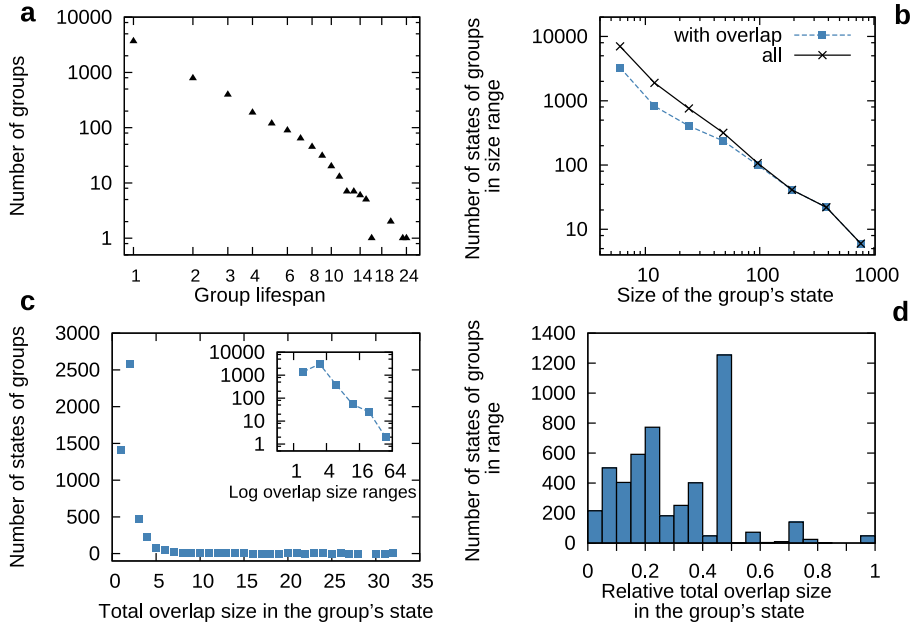}
  \caption{Statistical analysis of group lifespans and group overlaps.
    (a) Histogram of the groups' lifespans.
    (b) Size histogram for all states of the groups and for the states
    with overlap.
    (c) Total overlap size histogram of the groups that have at least $1$
    shared group member with another group.
    (d) Relative total overlap size histogram of the groups that have at least $1$
    shared group member with another group.
    As before, the obtained power law distributions in panels (a) and (b) 
    indicate that both group lifespans and group sizes are continuously 
    distributed between small and large values. 
    In other words, in either case one cannot select 
    a single set of large values and disregard all others.
}
  \label{fig5}
\end{figure*}

There are altogether $5,439$ groups made up by a total of $10,160$
yearly group states (``snapshots'' of timelines). 
Figure\,\ref{fig5}a shows the histogram of the groups' lifespans.
The majority of the groups exist only for 1 or 2 years 
that may be called ``rapid transient'' groups.
We found 43 groups with a lifespan of more than 10 years
and a maximum lifespan of 24 years (for one group).

The rapid transient groups show that in scientific research there is a
constant push for finding new solutions, i.e., launching new fields.
The main reasons for a group to be transient are the following:
\begin{itemize}
\item[(A)] The articles of a transient group discuss a current
  (usually technical) question, or support and validate a
  result. After resolving the issue, these articles are not co-cited
  any more. 
\item[(B)] Groups can merge: small transient groups form a single
  group. A typical case for this evolution pattern is the emergence of
  a new topic.
\item[(C)] Small transient groups provide a pool to absorb papers from
  for the more stable groups of ``canonical'' papers. In larger groups
  with a longer life span members do change, but within the group
  there is always a dominant subset that changes only slowly. As an
  example consider a group defined as the set of employees at a large
  company. Here usually middle managers and administrators provide the
  backbone of the company and represent most solidly its character
  (its company culture and core values).
\end{itemize}

Two identified groups of publications (articles)
overlap, if they share at least one article.
Articles in the overlap play a bridging role between the two groups.
They connect two separate topics or relate two scientific schools.
Note also that an overlap between two groups indicates a loose
connection between the entire groups including their further (not shared) members.
Groups without overlaps represent very special topics,
whereas groups dominated by shared nodes have little specificity.
Section\,\ref{clusters} explained how we applied the Clique Percolation Method (CPM, implemented by CFinder)
for identifying modules.
The CPM explicitly allows for overlaps among the modules.
The following paragraphs explain our analyses of the overlaps among the groups of co-cited publications.

Figure\,\ref{fig5}b compares 
(i) the sizes of those yearly co-cited article groups that have overlaps
and (ii) the sizes of all yearly states of groups.
The figure shows that the groups without overlap are usually smaller.
In the next two statistics we focus on the states with at least one
group member in an overlap.
For a selected group the total overlap size is
the number of the group's members that are at the same time
members of at least one other group in the same year.
Figure\,\ref{fig5}c shows the number of yearly 
co-cited article groups with a given total overlap size.
Similarly to the previous definition, for a selected group
the relative total overlap size is the total overlap size divided by the
number of items in group.
Figure\,\ref{fig5}d shows the histogram of the relative total overlap sizes.

In $97\%$ of the $4,921$ group states that have at least one shared (overlapping) member
the total overlap size is at most $5$ articles.
Moreover, based on the relative total overlap size 
in most cases ($94\%$) the relative total overlap is at most $0.5$.
States with size larger than $20$ articles have the relative total
overlap below $0.3$.
In summary, for the large yearly co-cited article groups
the relative amount of overlap is low,
however, for small groups overlaps can be important.

\begin{figure*}[h!]
  \includegraphics[width=1.00\textwidth]{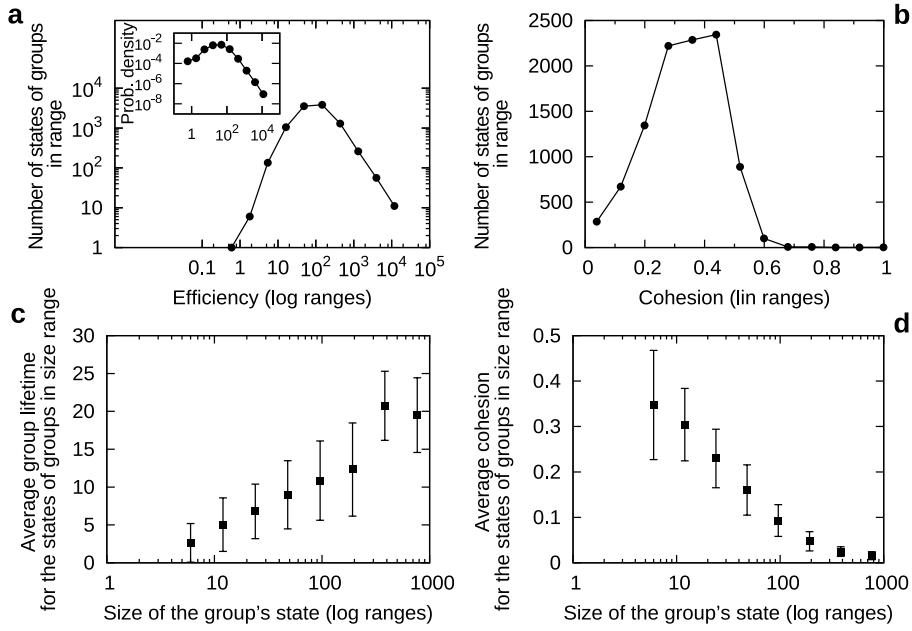}
  \caption{Statistical analysis of group efficiency and group cohesion.
    Note that in this figure the yearly snapshots of the co-cited groups of articles are called ``states of groups''.
    (a) Main panel: Histogram of group efficiency. Inset: Distribution (probability density function) of group efficiency.
        Group efficiency values are binned according to efficiency ranges increasing exponentially (i.e., the horizontal axis is logarithmic).
    (b) Histogram of group cohesion values.
        Group cohesions are binned. Bins have identical sizes (i.e., the horizontal axis is linear).
    (c) Average and standard deviation of the lifespan of groups as a function of group size.
        Group sizes are binned according to size ranges increasing exponentially (i.e., the horizontal axis is logarithmic).
    (d) Average and standard deviation of group cohesion as a function of group size.
        Group sizes are binned as in (c).
}
  \label{fig6}
\end{figure*}

Recall from Section\,\ref{subsec:influ}
that for the year $t$ the group efficiency, $\varepsilon(G(t))$, of group $G$
quantifies the effect that this group of publications has
on science in the year $t$ and short before that year.
According to Figure\,\ref{fig6}a, out of the major simple statistical distributions
this distribution of group efficiencies resembles a log-normal (or a power-law) distribution,
i.e., an upside down parabolic function (or a linear decay) on the log-log scale.
Both distributions (log-normal and power-law)
indicate that the total effect of a group of publications
arises not as a sum of independent effects,
but through reinforcing (e.g., multiplicative) effects.

Note that for almost $60\%$ of the yearly group states
the group efficiency is $\varepsilon \leq 100$.
On the other hand, there are three small group states with $\varepsilon > 10,000$.
They are due to three heavily cited articles \citep{bradford1976,sanger1977,chomczynski1987}.
In fact, in each of these three yearly group states
at least two of the three highly cited articles are present.
One of the three papers describes a method for protein identification,
the second describes a new method of total RNA isolation,
and the third presents method for determining nucleotide sequences in DNA.

As for group cohesion, recall from Section\,\ref{subsec:struct} that
$\kappa(G(t))$ is the number of citations within the article co-citation group $G(t)$ 
divided by the maximum possible number of citations among the articles of $G(t)$.
According to Figure\,\ref{fig6}b, for $\approx 83\%$ of the yearly group states
group cohesion is between $0.2$ and $0.5$.
To put this result in context, recall from Section\,\ref{data_methods}
that citations normally point from a later study to an earlier one.
In other words, group cohesion values above $0.5$ can appear only if 
there are mutual (bidirectional) citations between the group members.
Since mutual citations are rare, only around $1.4\%$ of the yearly states
of co-citation groups exceed the $0.5$ group cohesion value.
Finally, note in Figure\,\ref{fig6}c that a higher number of papers
in a co-citation group of papers usually implies that the given group
has a longer lifespan.

\subsection{Lifespans of the modules of co-cited papers and transitions between the modules}
\label{subsec:lifetrans}

Figure\,\ref{fig:timelines-size} summarizes the groups with at least $14$
years of lifespan and the articles transitioning between these groups.
%
%
Recall that in Figure\,\ref{fig:timelines-size} a large group of
papers transitioning from one field to another field
shows that the first field is likely to be in the process of
dissolving and the latter field is absorbing much of it. 
For our quantitative purposes, we define an article to be
transitioning if the article is a member of group $g_1$ at time $t$,
and it becomes a member of another group $g_2$ in the next time
step.
%
On this map note that groups usually start small, then they grow, and finally, they shrink before disappearing.
According to the most relevant WoS categories, indicated in front of
the groups, these longest lived groups are mostly related to biology.
Also, articles with the journal category
\textit{"multidisciplinary sciences"}
seem to be frequent in the majority of the groups.
Note also that
(i) topic tags may change if the article set included in the group changes
and (ii) the extracted topic set is also influenced by the paper
titles of the other groups in the same year via the Tf-Idf scoring.
The reason for (ii) is that the keywords present in many groups in a
particular year are considered to have less specificity
for characterizing the unique aspects of a group's state.

Next, we provide examples for published papers transitioning between co-citation groups.
Two groups with many transitions are No.5 (group G5) and No.7 (group G7).
Group G5 accumulates articles mostly related to the terms \textit{cell-death},
\textit{apoptosis}, \textit{tumours}, \textit{cancer}, \textit{cell-cycle}, and genes,
factors, proteins related to these topics.
The publications in Group G7 can be most closely characterized by \textit{gene},
\textit{protein}, \textit{dna} and \textit{sequences}.
The lifespan of this group (G7) can be partitioned into three time ranges. 
For each of these three time ranges we mention characteristic topics here.
The beginning of the first time range is dominated by papers focusing on the investigation of \textit{simian-virus-40}.
Then, this time range can be characterized by the topics:
\textit{gene}, \textit{dna}, \textit{sequence} and it is also related
to \textit{viruses}.
In the second time range \textit{gene} is still important and there is
orientation towards \textit{oncogene}, \textit{proto-oncogene}, \textit{tumour}, \textit{cancer}.
In the last range, many publications are related to
\textit{protein}, \textit{receptor}, \textit{kinase}, \textit{activation} or \textit{activity}.

From 1993 to 1994, a set of 107 articles left the timeline 
(group) G7 and became members of timeline (group) G5.
These articles were mostly related to the following topics: \textit{receptor},
\textit{sequence}, \textit{gene}, \textit{dna}.
In 1994 another large group of papers left G5.
From 1994 to 1995 the timeline of G5 lost 56 articles to G2, the main topics of these articles were
\textit{sequence}, \textit{dna}, \textit{rna} and \textit{protein}.
These transitioning articles are all related to
either the gene JUN or the transcription factor AP-1.
The gene JUN participates in cell cycle progression control and regulates apoptosis \citep{wisdom_etal1999}.
Observe that timeline G2 started with 5 articles in 1994,
of which 4 had been the members of G7 in the previous year (1993).
To summarize the early years of group G2,
the \mbox{G7,1993 $\rightarrow$ G2,1994} transition contributes to the birth of group G2,
while the transition \mbox{G5,1994 $\rightarrow$ G2,1995} shifts the group's 
topic profile towards molecular biological methods.
Observe in Figure\,\ref{fig:timelines-size} 
that between 1991 and 1992 another 16 articles moved from G6 to G9.
This shift includes articles with the topic \textit{endothelin}, and largely contributes
to the topic of group G9, which is related to \textit{endothelin}.

Finally, we discuss a case when the topic set of past papers changes as these papers move from 
one group to another.
In Figure\,\ref{fig:timelines-size} consider the transition 
(G7, 1993, [receptor;activ;kinas]) $\rightarrow$ (G5, 1994, [transcript;mutat;p53]).
The starting point of this transition is the 1993 snapshot of group G7
with the main topics ``receptor'', ``activ'' and ``kinas''.
By the next year (1994) many of the papers from this densely co-cited group move over to G5.
To analyze this transition at a higher resolution, we now consider now one of the highly cited papers participating in this transition \citep{altschul1990}.
The article by Altschul et. al. introduced BLAST, a new approach to rapid sequence
comparison based on local sequence similarity.
Since its publication, BLAST has become widely used for DNA and protein sequence database searches.
Table\,\ref{tab:blast} shows the groups which this paper participated in.
First, it appeared in group G7 in 1992, and remained in the same group in 1993.
After its transition to G5 in 1994, the BLAST paper moved to G2 in 1995 and remained in this group until 2008.

\begin{table}[!ht]
\caption{
Membership of the BLAST article \citep{altschul1990} in the timelines constructed from 
yearly co-citation groups of publications.
For each year the topic tag set of the article's group is shown below.
Groups G2, G7 and G5 are displayed on the map of Figure\,\ref{fig:timelines-size},
while g1, g2 and g3 are small and short-lived groups that are mentioned in this table only because
they contain the BLAST article as well.
Note that due to group overlaps a single paper can be contained by more than one group at the same time,
for example, in the year 2000 the BLAST paper is a member of groups G2 and g3.
}
\label{tab:blast}
\begin{tabular}{ll}
\hline\noalign{\smallskip}
Year & Timeline: tag set\\
\noalign{\smallskip}\hline\noalign{\smallskip}
1992 & \textbf{G7}: retino acid;cystic-fibrosi;element\\
1993 & \textbf{G7}: receptor;activ;kinas\\
\noalign{\smallskip}\hline\noalign{\smallskip}
1994 & \textbf{G5}: transcript;mutat;p53\\
\noalign{\smallskip}\hline\noalign{\smallskip}
1995 & \textbf{G2}: sequenc;dna;method \& \textbf{g1}: human brain;cdna sequenc;express sequenc tag\\
1996 & \textbf{G2}: sequenc;align;dna\\
1997 & \textbf{G2}: rapid;rna-polymeras;method\\
1998 & \textbf{G2}: sequenc;tree;method\\
1999 & \textbf{G2}: sequenc;tree;method \& \textbf{g2}: search;protein\\
2000 & \textbf{G2}: sequenc;phylogenet;method \& \textbf{g3}: search;autom sequenc;trace\\
2001 & \textbf{G2}: protein;program;structur\\
2002 & \textbf{G2}: protein;program;method\\
2003 & \textbf{G2}: molecular;method;genom\\
2004 & \textbf{G2}: protein;molecular;network\\
2005 & \textbf{G2}: model;molecular;method\\
2006 & \textbf{G2}: model;molecular;method\\
2007 & \textbf{G2}: model;molecular;method\\
2008 & \textbf{G2}: model;calcul;method\\
\noalign{\smallskip}\hline
\end{tabular}
\end{table}

\subsection{Multidisciplinary papers frequently form strongly co-cited groups}
\label{multigroups}

\begin{figure*}[h!]
  \includegraphics[width=1.1\textwidth]{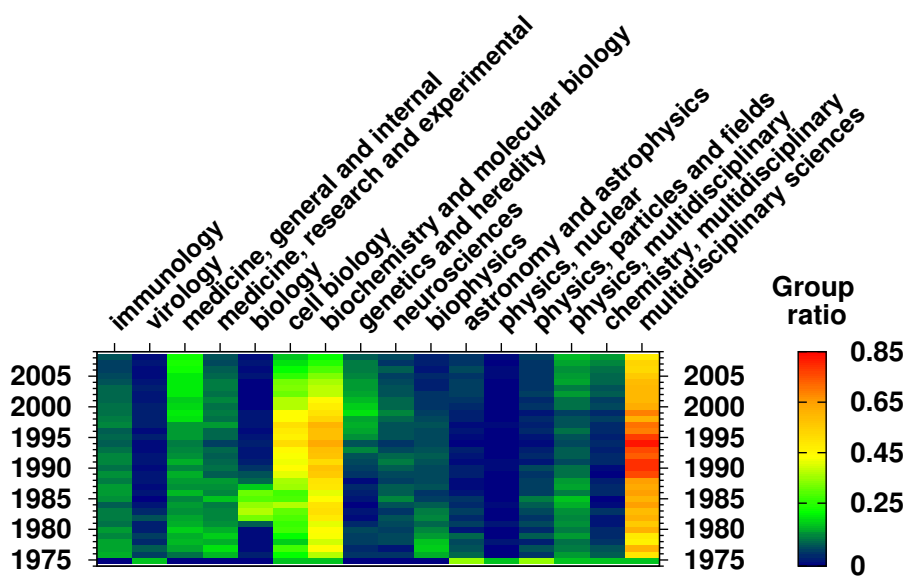}
  \caption{Presence of the Web of Science categories in each year's co-citation groups.
    For each year and each category the ``Group ratio'' 
    (see colour scale on the right) is the number of groups with the 
    selected category in the given year divided by the number of 
    all groups in that year.
    %
    %
    (Categories assigned to more than $10\%$ of the group member
    articles in a certain year are assigned to the group in that
    year.
    Only categories assigned to more than $10\%$ of the groups in at least
    one year from 1975 to 2008 are shown.)
    Observe that co-citation groups are enriched with 
    multidisciplinary papers (compare also to\,\ref{fig3}). 
    To the best of our knowledge even those who consider this
    statement trivial have not yet quantified it at the large
    scale and detail applied in the current paper.}
  \label{fig8}
\end{figure*}

According to Figure\,\ref{fig8}, in almost every year
\textit{"multidisciplinary sciences"}, 
\textit{"biochemistry and molecular biology"} and
\textit{"cell biology"} are over-represented in the groups
of co-cited publications.
Of these three,
\textit{"multidisciplinary sciences"} is even more outstanding:
between 1989 and 1995 it is associated with
more than $70\%$ of the groups in each year.
As a contrast, Figure\,\ref{fig3} shows that every year the category
\textit{"multidisciplinary sciences"} appears on not more than 
approximately $5\%$ of all published articles.

\newpage

\begin{figure*}[h]
  \begin{center}
  \includegraphics[width=0.7\textwidth]{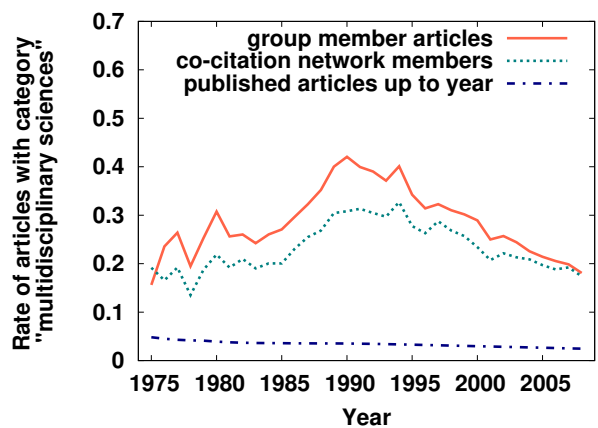}
  \caption{Annual multidisciplinarity ratio among the co-citation
    group members, co-citation network members and 
    among all articles published up to the given year.
    As in Figure\,\ref{fig8}, the enrichment of multidisciplinary papers is clear
    when moving from the set of all papers to co-cited papers 
    and then to co-citation groups.}
  \label{fig9}
  \end{center}
\end{figure*}

\noindent
Triggered by this observation we analyzed in more detail the usage of the 
category ``multidisciplinary sciences''.
This category contains articles that were published in journals of a very general character.
We calculated the ratio of ``multidisciplinary sciences''
articles for each year in the following article sets:
(i) published articles up to the given year,
(ii) papers co-cited in the given year (with the optimized link weight threshold, see Section\,\ref{clusters}), and the
(iii) papers belonging to the co-citation groups of the given year.
Here we excluded articles without category information.
From this point on, for the sake of simplicity the 
fraction of all papers labelled with the category ``multidisciplinary sciences''
will be called the {\bf multidisciplinarity ratio}.
According to Figure\,\ref{fig9}, the multidisciplinarity ratio is
significantly higher in the co-citation networks than among all published articles.
Moreover, further enrichment is observed in the co-citation network groups.

\begin{figure*}[h]
  \begin{center}
  \includegraphics[width=0.7\textwidth]{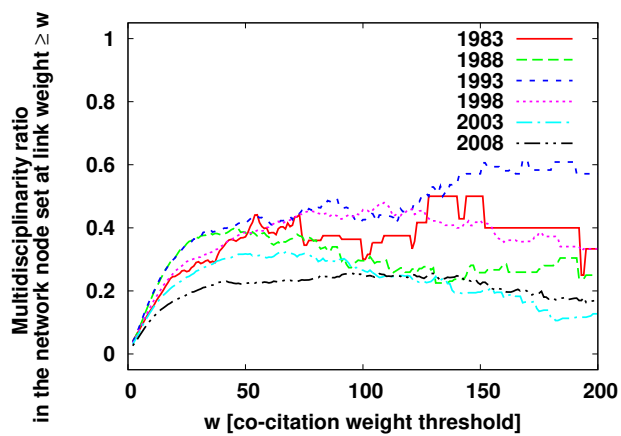}
  \caption{Multidisciplinarity ratio in the network when
    increasing the co-citation link weight threshold.}
  \label{fig10}
  \end{center}
\end{figure*}

We find that -- compared to all published articles --
the co-citation networks (at the optimized link weight threshold)
and the groups of co-cited articles 
are enriched for articles that are multidisciplinary.
Following this observation, our next goal was to measure 
how the co-citation link weight threshold influences
this enrichment.
Therefore, for each publication year we scanned through all possible co-citation
network weight thresholds and calculated the multidisciplinarity ratio
for those network nodes that have at least one link not weaker than the given threshold.
We found that for low link weight thresholds the multidisciplinarity ratio tends to grow as the threshold grows.
The range where this relationship holds may differ for the individual years.
We show examples for this initial range in Figure\,\ref{fig10}.
For instance, in 1983 the multidisciplinarity ratio grows as a function of the link weight threshold ($w$)
up to approximately $w=55$.
We note that for each year the optimized weight threshold used for identifying groups in Section\,\ref{clusters} falls into this special initial range of that year.

\newpage
\section{Conclusions}
\label{conclusion}
We identified and analyzed the dense cores of yearly article
co-citation networks based on Web of Science citations (1975-2008).
We joined the identified yearly co-citation paper clusters (also called: groups, modules or intellectual bases) to obtain the timelines of scientific fields.
Next, we quantified the structure of co-citation groups (through
group lifespan, size, efficiency, cohesion, and overlap)
and listed the topics of the groups (through title-based tags and WoS category terms).
We applied a task-specific combination of methods:
(i) the Clique Percolation Method and the module joining method for identifying the groups of articles and the timelines,
and (ii) term frequencies, Tf-Idf scores, and the rCUR dimension reduction technique to find appropriate tags for the groups.

We found that most groups have short lifespans (1 or 2 years) and that larger groups tend to live longer.
Among the few long-lived groups the longest lifespan is 24 years.
This group (G5 on Figure\,\ref{fig:timelines-size}) focuses on the topics
\textit{cell-death}, \textit{apoptosis}, \textit{tumours}, \textit{cancer} and \textit{cell-cycle}.
We found that for large groups the total relative overlap (the ratio of a group's nodes that it shares with other groups) is low,
but for small groups it can be high.
We quantified the influence of each group on science over time by calculating its yearly efficiency 
(the efficiency measures the number of articles citing the group from outside).
A few of the groups have high efficiencies, which is usually caused by a few of their highly cited articles.
However, the efficiency of most groups in most years is much lower than these top values.

We constructed a map of the co-cited groups with the longest lifespans (Figure\,\ref{fig:timelines-size}).
This map includes the groups' most relevant Web of Science (WoS)
categories, group sizes, characteristic topics and article transitions
between the groups.
This map helped us to track in time the groups with high
group transition activity, and to survey the topic shifts of the groups.
Moreover, we analyzed in detail one selected article on which topic labels changed over time.
Also, we noticed that the WoS category \textit{``multidisciplinary sciences''}
is over-represented in the co-cited article groups in almost every year
as compared to all published papers.
We found that the cores of the co-cited articles are mainly multidisciplinary.

To the best of our knowledge the paper \citep{sinatra2015}
is one of the very few that are similar both in their topic and their
scale to the current paper.
Sinatra et. al. have quantified the ``rapid growth and increasing
multidisciplinarity of physics'' in groups of publications defined by
scientific literature classification schemes.
As a contrast, 
\begin{itemize}
\item[(1)] the current paper works with the entire Web of Science,
\item[(2)] when it discusses multidisciplinarity, it emphasizes not
  growth, but relative amount, and 
\item[(3)] it quantifies multidisciplinarity in groups of papers
  identified with a special novel combination of numerical
  techniques. 
\end{itemize}

Regarding further challenges and the limitations of the approach explained in the current paper, first note that we used the time dependent version of the Clique Percolation Method, which focuses on the most connected parts of the co-citation network.
Hence, several papers that are cited regularly, but not always co-cited with the same papers, are out of the scope of this work. 
Second, co-citation does not always indicate close relatedness.
For example, most scientific publications start with an introductory part in which the authors put their work into a wide context, thus, the papers cited in an introduction can have very different topics.
Third, in any field of science citations are not equal:
a handful of citations may be crucial while others remain less important.
The most straightforward way to account for this effect is to construct the yearly co-citation networks by applying citations that are weighted by relevance, e.g., their directed edge betweenness centralities.

Finally, we note that from the point of view of earlier research
\citep{levitt_and_thelwall2008} it is unexpected that
multidisciplinary papers are co-cited for a long time. 
Such stability was earlier thought to be characteristic of monodisciplinary papers only due to their clearly defined topic.
Note also that the current paper's results go beyond co-citation networks.
Compared to the co-citation network we find an even higher concentration of multidisciplinarity in the evolving dense groups (modules) of co-cited papers.

\begin{acknowledgements}
We thank Tam\'{a}s Vicsek, Gergely Palla and B\'{a}lint T\'{o}th for discussions and advice.
This project was supported by the European Union
and the European Social Fund through the FuturICT.hu project
(Grant ID: TAMOP-4.2.2.C-11/1/KONV-2012-0013)
and the Hungarian National Science Fund (Grant ID: OTKA K105447).
\end{acknowledgements}

\clearpage
\bibliographystyle{spbasic}      
\bibliography{bib.bib}   

%
%

\end{document}